\documentclass[conference]{IEEEtran}
\IEEEoverridecommandlockouts
\usepackage{cite}
\usepackage{amsmath,amssymb,amsfonts,amsthm,bbm}
\usepackage{stmaryrd}
\usepackage{graphicx}
\usepackage{textcomp}
\usepackage{xcolor}
\usepackage{booktabs}
\usepackage{caption}
\usepackage{subcaption}
\usepackage{optidef}

\usepackage{makecell}
\usepackage{multirow}
\usepackage{arydshln}
\usepackage[all=normal,paragraphs=tight,floats=tight,mathspacing=normal,wordspacing=tight,charwidths=tight,mathdisplays=normal,leading=tight]{savetrees}

\newcommand{\abs}[1]{\ensuremath{\left| #1 \right|}}

\DeclareMathOperator*{\argmax}{arg\,max}

\newcommand{\norm}[2]{\ensuremath{\left\lVert #1 \right\rVert}_#2}

\newcommand*{\herm}{{\mathsf{H}}}

\usepackage{algorithm}
\usepackage{algpseudocode}

\algnewcommand\algorithmicforeach{\textbf{for each}}
\algdef{S}[FOR]{ForEach}[1]{\algorithmicforeach\ #1\ \algorithmicdo}

\def\BibTeX{{\rm B\kern-.05em{\sc i\kern-.025em b}\kern-.08em
    T\kern-.1667em\lower.7ex\hbox{E}\kern-.125emX}}

\begin{document}
\bstctlcite{IEEEtran:BSTcontrol}

\title{CSI Compression using Channel Charting}

\author{
	\IEEEauthorblockN{
        Baptiste Chatelier\IEEEauthorrefmark{3}$^,$\IEEEauthorrefmark{2}$^{,\star}$,
        Vincent Corlay\IEEEauthorrefmark{3}$^{,\star}$,
        Matthieu Crussière\IEEEauthorrefmark{2}$^{,\star}$,
		Luc Le Magoarou\IEEEauthorrefmark{2}$^{,\star}$
		}
	\IEEEauthorblockA{
		\IEEEauthorrefmark{2}Univ Rennes, INSA Rennes, CNRS, IETR-UMR 6164, Rennes, France
	}
	\IEEEauthorblockA{
		\IEEEauthorrefmark{3}Mitsubishi Electric R\&D Centre Europe, Rennes, France
	}
	\IEEEauthorblockA{
		$^\star$b\raisebox{0.2mm}{\scalebox{0.7}{\textbf{$<>$}}}com, Rennes, France
	}
	}
\maketitle

\begin{abstract}
    Reaping the benefits of multi-antenna communication systems in frequency division duplex (FDD) requires channel state information (CSI) reporting from mobile users to the base station (BS). Over the last decades, the amount of CSI to be collected has become very challenging owing to the dramatic increase of the number of antennas at BSs. To mitigate the overhead associated with CSI reporting, compressed CSI techniques have been proposed with the idea of recovering the original CSI at the BS from its compressed version sent by the mobile users. Channel charting is an unsupervised dimensionality reduction method that consists in building a radio-environment map from CSIs. Such a method can be considered in the context of the CSI compression problem, since a chart location is, by definition, a low-dimensional representation of the CSI. In this paper, the performance of channel charting for a task-based CSI compression application is studied. A comparison of the proposed method against baselines on realistic synthetic data is proposed, showing promising results.
\end{abstract}


\IEEEpeerreviewmaketitle

\section{Introduction}\label{sec:introduction}
\IEEEPARstart{T}{he} shift towards higher operating frequencies in the Fifth Generation (5G) and Beyond 5G (B5G) communication systems has led to the widespread use of antenna arrays made up of a large number of smaller radiating elements, capable of forming high-gain directional beams. However, this increase in the BSs antenna number causes a significant overhead in the Channel State Information (CSI) reporting procedures in Frequency Division Duplex (FDD) schemes. Indeed, in FDD systems, in order to perform downlink channel estimation, User Equipments (UEs) need to periodically report one complex number per antenna and subcarrier to the BS. To mitigate this reporting overhead, CSI compression techniques have been envisioned~\cite{ChaoKai18}. The associated reporting procedure can be summarized as follows: in a FDD system, a BS sends pilot symbols to a UE that uses them to perform CSI estimation. The UE then compresses the estimated CSI and transmits it to the BS. Leveraging its readily available computing power, the BS is then able to reconstruct the CSI with minimal distortion, before using it to perform communication tasks such as beamforming or resource allocation. This problem has been extensively studied in the past few years, with Artificial Intelligence/Machine Learning (AI/ML) methods proposed to address it. A non-exhaustive list of relevant studies includes~\cite{ChaoKai18,Jiajia20,Wang19,Lu20,Chen22,Cui22}. An in-depth review of the deep learning based CSI compression literature is provided in~\cite{Guo22}.

This paper considers the task-based CSI compression problem from a different perspective, using channel charting (CC). Since the reconstructed CSI is primarily used for beamforming, one could consider the CSI compression problem where the decoding step reconstructs a precoding matrix rather than the CSI. Channel charting~\cite{Studer18} is an unsupervised dimensionality reduction methods that compress CSI into chart locations. More precisely, the goal of CC is to embed high dimensional CSIs into a low dimensional space that preserves local neighborhoods: channels from close locations should be embedded into chart locations that are close one to each-other. CC has been used for beam prediction~\cite{Ponnada21,Yassine23}, precoder learning~\cite{LeMagoarou22}, but also for positioning~\cite{Euchner23,Taner24}. In the context of CSI compression, it clearly appears that CC can be seen as a good candidate for the encoding stage, as, by definition, a chart location is a compressed version of its associated CSI.

\noindent\textbf{Contributions.} This paper considers the concept of task-based CSI compression: its goal is to reduce the CSI reporting overhead while optimizing beamforming performance. The encoder consists of a CC model, while the decoder is a neural network. An end-to-end learning strategy for the encoder and decoder training is presented, with a task-oriented loss function. Furthermore, the number of learnable parameters is optimized through a subsampling approach within the encoder, resulting in a ninetyfold reduction in the number of learnable parameters, without significant performance degradation. Finally, simulations on realistic data show that the proposed approach achieves excellent beamforming performance with significantly higher compression ratios (around 1024) compared to those obtained by classical auto-encoders (around 64).

\noindent\textbf{Related work.} Inferring a precoder from a chart location has been studied for Cell Free massive MIMO systems in~\cite{LLM22}. However, that study did not optimize the encoder complexity and only trained the decoder network. Similar studies have been carried out with precoders constrained in a codebook in~\cite{Ponnada21,Yassine23}. Jointly optimizing CSI compression and beamforming performance has been studied in~\cite{Guo21,Sun22,Qiulin23}. However, none of these studies considered channel charting for the encoding stage. Finally, CSI compression using AI/ML methods have been extensively studied in~\cite{ChaoKai18,Jiajia20,Wang19,Lu20,Chen22,Cui22}. Such approaches obtain good performance in CSI reconstruction but often present poor compression ratio compared to our approach as they are not specialized to a given task.

\section{Problem formulation}\label{sec:problem_formulation}

This paper considers a multi-user Multiple Input Multiple Output (MU-MIMO) system comprised of a BS equipped with $N_a$ antennas and $N_u$ mono-antenna UEs. The BS transmits symbols to the UEs on $N_s$ subcarriers.

% \begin{figure}[!h]
% 	\centering
% 	\includegraphics[scale=.5]{figs/system_model.pdf}
% 	\caption{System model and reporting procedure}
% 	\label{fig:system_model}
% \end{figure}

% \noindent\textbf{Reporting procedure.} Fig.~\ref{fig:system_model} depicts the system model as well as the reporting procedure. A BS sends pilot symbols to a set of UEs. The UEs use the pilots to estimate their downlink channel $\mathbf{h}_k \in \mathbb{C}^D$. Note that vectorized channels are considered, so that $D=N_a N_s$. Then, UEs perform CSI compression: the downlink channel is encoded into a low dimensional vector $\mathbf{z}_k \in \mathbb{R}^d$. Finally, the BS receives the encoded channels from all UEs and decode them into a precoding matrix $\mathbf{W}$.
\noindent\textbf{Reporting procedure.} A BS sends pilot symbols to a set of UEs. The UEs use the pilots to estimate their downlink channel $\mathbf{h}_k \in \mathbb{C}^D$. Note that vectorized channels are considered, so that $D=N_a N_s$. Then, UEs perform CSI compression: the downlink channel is encoded into a low dimensional vector $\mathbf{z}_k \in \mathbb{R}^d$. Finally, the BS receives the encoded channels from all UEs and decode them into a precoding matrix $\mathbf{W}$, used to perform spatial multiplexing.

\noindent\textbf{Objective.} The goal of this paper is to design an efficient CSI compression method that solves the following sum-rate (SR) optimization problem in a parallel transmission to $K \leq N_u$ UEs:
\begin{maxi}|s|
	{\mathbf{W} \in \mathbb{C}^{D\times K}}{\sum_{k=1}^K \log_2 \left(1+\dfrac{\abs{\mathbf{h}_k^\herm \mathbf{w}_k}^2}{\sigma_k^2 + \sum\limits_{j\neq k} \abs{\mathbf{h}_k^\herm \mathbf{w}_j}^2}\right),}
	{}{}
	\addConstraint{\mathcal{C}_{p}: \forall k \in \llbracket 1, K \rrbracket, \norm{\mathbf{w}_k}{2} = \dfrac{1}{\sqrt{K}}}{}
\end{maxi}
where $\mathbf{w}_k$ is a precoding vector decoded from a compressed representation of the channel $\mathbf{h}_k$ and $\sigma^2_k$ is the associated noise variance. The power constraint $\mathcal{C}_{p}$ considers uniform power allocation between each UE.
% In a parallel transmission to $K \leq N_u$ UEs, the received signal can be defined as:
% \begin{equation}
% 	\mathbf{y} = \mathbf{H}^\herm \mathbf{Ws}+\mathbf{n}
% \end{equation}
% with $\mathbf{y} \in \mathbb{C}^{K}$ being the received signal, $\mathbf{H} = \{\mathbf{h}_k\}_{k=1}^K \in \mathbb{C}^{D \times K}$ the channel matrix, $\mathbf{W} = \left\{ \mathbf{w}_k\right\}_{k=1}^K \in \mathbb{C}^{D \times K}$ the beamforming matrix, $\mathbf{s} \in \mathbb{C}^{K}$ the transmitted symbols, and $\mathbf{n} \in \mathbb{C}^{K} \sim \mathcal{CN}\left(\mathbf{0}_K,\sigma^2_n\mathrm{Id}_K\right)$ the transmission noise. The goal of this paper is to design an efficient CSI compression method that solves the following sum-rate (SR) optimization problem:
% \begin{maxi}|s|
% 	{\mathbf{W}}{\sum_{k=1}^K \log_2 \left(1+\dfrac{\abs{\mathbf{h}_k^\herm \mathbf{w}_k}^2}{\sigma_k^2 + \sum\limits_{j\neq k} \abs{\mathbf{h}_k^\herm \mathbf{w}_j}^2}\right)}
% 	{}{}
% 	\addConstraint{\mathcal{C}_{p}: \forall k \in \llbracket 1, K \rrbracket, \norm{\mathbf{w}_k}{2} = \dfrac{1}{\sqrt{K}}}{}
% \end{maxi}
% where $\mathbf{w}_k$ is a precoding vector decoded from a compressed channel. The power constraint $\mathcal{C}_{p}$ considers uniform power allocation between every UE.

The foundation of the proposed method is formally based on defining a well suited encoding function $\mathcal{E}$ and decoding function $\mathcal{D}$. The role of the encoding function is to project the CSI into a low-dimensional space:
\begin{equation}
	\begin{aligned}
		\mathcal{E}\colon \mathbb{C}^D &\longrightarrow \mathbb{R}^d\\
		\mathbf{h} &\longrightarrow \mathbf{z} \triangleq \mathcal{E}\left(\mathbf{h}\right),
	\end{aligned}
\end{equation}
where $d \ll D$. Conversely, the decoding function maps the low-dimensional channel charts to a precoding matrix:
\begin{equation}
	\begin{aligned}
		\mathcal{D}\colon \mathbb{R}^d \times \cdots \times \mathbb{R}^d &\longrightarrow \mathbb{C}^{D \times K}\\
		% \mathcal{D}\colon \left(\mathbb{R}^d\right)^K &\longrightarrow \mathbb{C}^{D \times K}\\
		\left\{\mathbf{z}_k\right\}_{k=1}^K &\longrightarrow \mathbf{W} \triangleq \mathcal{D}\left(\left\{\mathbf{z}_k\right\}_{k=1}^K\right).
	\end{aligned}
\end{equation}

\noindent \textbf{Performance measure.} The performance of the proposed method is measured in different configurations: mono and multi-UE scenarios. In a mono-UE scenario, it has been shown in~\cite{LeMagoarou22} that the transmission capacity is linked to the squared cosine similarity between the channel $\mathbf{h}_k$ and the associated inferred precoder $\mathbf{v}_k$:
\begin{equation}
	\rho_k = \dfrac{\abs{\mathbf{v}_k^\herm \mathbf{h}_k}^2}{\norm{\mathbf{v}_k}{2}^2\norm{\mathbf{h}_k}{2}^2}.
\end{equation}
This metric is normalized within the $\left[0,1\right]$ range. $0$ means that the precoder is orthogonal to the channel, leading to no transmission, while $1$ means that the precoder is perfectly collinear with the channel. The mean and median correlations over the UE set will be considered.

In a multi-UE scenario, the performance measure is the sum-rate averaged across distinct groups of UEs in the location space. Considering the partition of the UE set $\mathcal{T}_u$ into $B$ groups of $K$ UEs: $\mathcal{T}_u = \bigcup_{b=1}^{B} \mathcal{T}_{u,b}$ with $\abs{\mathcal{T}_{u,b}} = K$ and $\forall m \neq n$, $\mathcal{T}_{u,m} \cap \mathcal{T}_{u,n} = \varnothing$, the sum-rate can be defined as:

\begin{equation}
	\overset{\circ}{\mathcal{R}} = \dfrac{1}{B} \sum_{b=1}^B \sum_{k} \log_2\left(1+\dfrac{\abs{\mathbf{h}_{k,b}^\herm \mathbf{w}_{k,b}}^2}{\sigma_{k,b}^2 + \sum\limits_{j\neq k} \abs{\mathbf{h}_{k,b}^\herm \mathbf{w}_{j,b}}^2}\right),
	\label{eq:erg_sr}
\end{equation}
where $\mathbf{h}_{i,b} \in \mathcal{T}_{u,b}$ if the channel for the $i$th UE of the $b$th group and $\mathbf{w}_{i,b}$ is the associated decoded precoder.

\section{Proposed approach}\label{sec:proposed_approach}
\begin{figure}[t]
	\centering
	\includegraphics[width=\columnwidth]{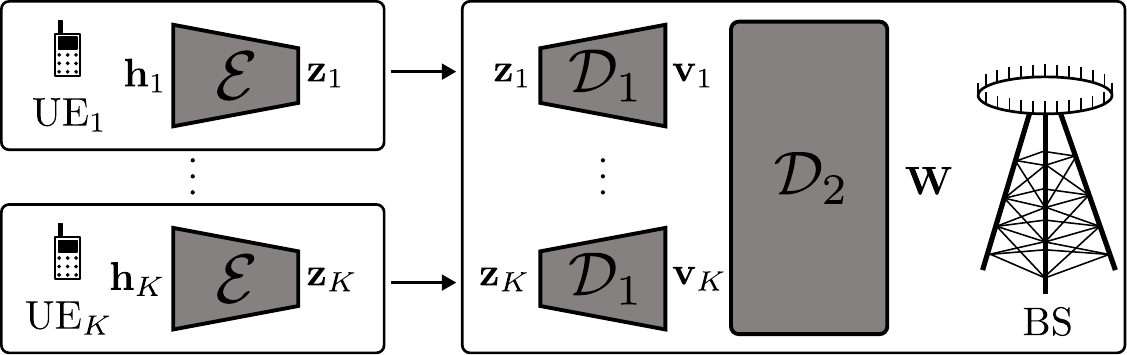}
	\caption{Overview of the proposed approach: encoder (left), decoder (right)}
	\label{fig:enc_dec_overview}
\end{figure}

This section presents the chosen models for the encoding and decoding functions, as well as the learning strategy and the learnable parameter reduction method. The CC-based encoder and decoder models are similar to those proposed in~\cite{LLM22,Yassine22,Yassine23}. An overview of the proposed approach is presented in Fig~\ref{fig:enc_dec_overview}.

\begin{figure}[b]
	\centering
	\includegraphics[width=\columnwidth]{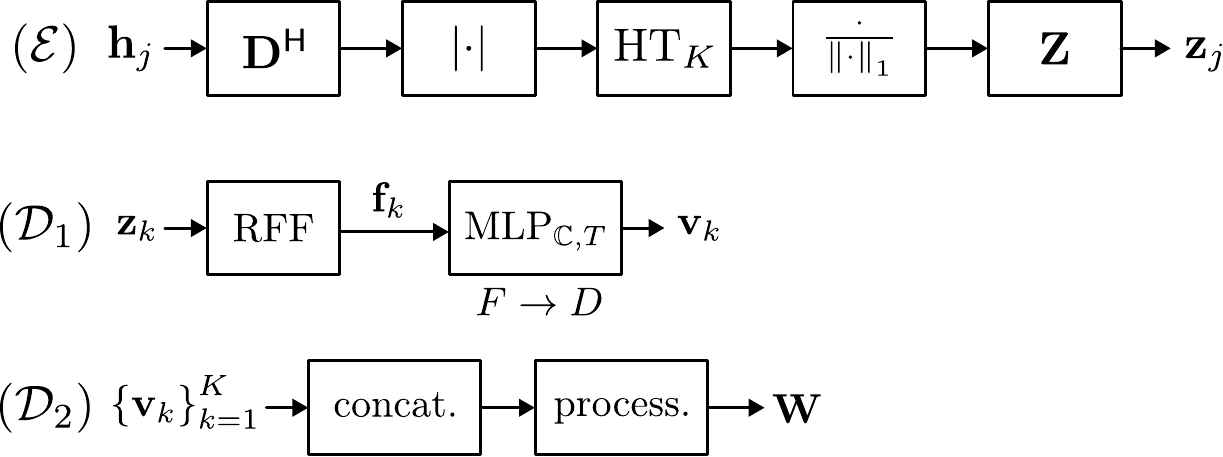}
	\caption{Encoder~\cite{LLM22,Yassine22} and decoder architectures}
	\label{fig:enc_dec_models}
\end{figure}

\noindent\textbf{Encoder.} As in~\cite{LLM22,Yassine22,Yassine23}, the encoding function is done through CC with a neural model presented in Fig.~\ref{fig:enc_dec_models}. This model is initialized through matrices $\mathbf{D}$ and $\mathbf{Z}$. Matrix $\mathbf{D} = \left\{ \mathbf{h}_i \right\} \in \mathbb{C}^{D\times N_c}$ consists of the concatenation of $N_c$ calibration channel vectors, while $\mathbf{Z}$ is initialized through the \texttt{ISOMAP} algorithm. This algorithm embeds the calibration channels into chart locations $\mathbf{z}_k \in \mathbb{R}^d$ ($d\ll D$). Note that the phase-insensitive distance of~\cite{LeMagoarou21} is used in the \texttt{ISOMAP} algorithm. The chart locations are then concatenated into matrix $\mathbf{Z} = \left\{\mathbf{z}_i\right\}_{i=1}^{N_\text{c}} \in \mathbb{R}^{d \times N_c}$. As many other dimensionality reduction methods, \texttt{ISOMAP} is known to be computationally intensive in out-of-sample scenarios, which reduces the efficiency of the encoding stage for channels not included in the calibration set. The presented CC model aims to address this issue by mapping an unseen channel $\mathbf{h}_j$ into a wisely chosen convex combination of calibration chart locations.

% \begin{figure}[b]
% 	\centering
% 	\includegraphics[width=\columnwidth]{figs/encoder.pdf}
% 	\caption{Encoder architecture~\cite{LLM22,Yassine22}}
% 	\label{fig:cc_model}
% \end{figure}

% \begin{figure}[b]
% 	\centering
% 	\includegraphics[scale=.45]{figs/decoder.pdf}
% 	\caption{Decoder architecture}
% 	\label{fig:decoder}
% \end{figure}

\noindent\textbf{Decoder.} It is proposed to split the decoder into two parts $\mathcal{D}_1$ and $\mathcal{D}_2$, so that $\mathcal{D} = \mathcal{D}_2 \circ \mathcal{D}_1$, as presented in Fig.~\ref{fig:enc_dec_models}. The first stage, $\mathcal{D}_1$, allows to map the chart locations to mono-UE precoding vectors. Then, the second stage, $\mathcal{D}_2$, is designed to concatenate and process the mono-UE precoding vectors into a multi-UE precoding matrix. The mapping of chart locations to precoding vectors is done through a neural network. Note that such mapping can present relatively fast variations in the embedding space. Indeed, the optimal precoders may vary rapidly in some regions of the location space. As channels originating from nearby locations are expected to be embedded to close chart locations through CC, this fast variation behavior is expected to be translated in the embedding space. As a result, it is proposed to use a random Fourier features (RFF)~\cite{Rahimi07} network for the chart-location-to-precoder mapping learning. Indeed, classical neural architectures such as MLPs have been shown to be biased towards learning low frequency functions~\cite{Rahaman2019,ijcai2021p304,JohnXu2020}: RFF networks have been proposed as a solution to this spectral bias issue~\cite{tancik2020fourfeat}. The \texttt{RFF} block is defined as:
\begin{equation}
	\mathbf{f}_k = \begin{bmatrix}
		\cos\left(2\pi\mathbf{B}\mathbf{z}_k\right)\\
		\sin\left(2\pi\mathbf{B}\mathbf{z}_k\right)
	\end{bmatrix},
\end{equation}
where $\mathbf{B} \in \mathbb{C}^{F\times d}$ is a frequency matrix initialized as $ \mathbf{B} \sim \mathcal{N}\left(\mathbf{0}_F,\sigma^2_B \mathrm{Id}_F\right)$. Note that $\sigma_B$ is an hyperparameter linked to the variation intensity in the chart location space: a higher $\sigma_B$ facilitates the decoder network's ability to learn rapid variations in the target mapping. The $\texttt{MLP}_{\mathbb{C},T}$ block denotes a three-layer MLP of size $T$, with $\texttt{ReLU}_{\mathbb{C}}$\footnote{Note that $\forall z_1 \in \mathbb{C}, \texttt{ReLU}_{\mathbb{C}}\left(z_1\right) = \texttt{ReLU}\left(\Re \left(z_1\right)\right) + \mathrm{j} \texttt{ReLU}\left(\Im \left(z_1\right)\right)$.} activation functions and complex weights and biases. 

The second decoding stage consists of a concatenation and a processing block. The role of the processing block is to introduce inter-UE interference cancellation in the concatenated channel matrix. This could be achieved using a neural network; however, the proposed approach leverages classical linear precoding techniques such as Maximum Ratio Transmitting (MRT) or Minimum Mean Squared Error (MMSE). These methods are grounded in well-established statistical criteria, offering robustness and interpretability. Thus, the processing block of $\mathcal{D}_2$ only consists of a denormalization step, linear matrix operations, and a normalization step to respect the uniform power allocation constraint. A more comprehensive discussion on this matter is presented below.

\noindent\textbf{Learning strategy.} It is proposed to train the proposed model using an end-to-end approach and a task-oriented loss function. As the objective is to maximize the sum-rate, one could use the negative sum-rate of Eq.~\eqref{eq:erg_sr} as a loss function. However, Eq.~\eqref{eq:erg_sr} depends on the UEs noise variance and on the number of parallel UEs. This implies that, when using Eq.~\eqref{eq:erg_sr} as a loss function, the trained encoder and decoder would be optimal only for a fixed noise variance and a fixed number of parallel UEs, limiting the practical interest of the proposed method. It is suggested to use a heuristic learning strategy that considers the following loss function:
\begin{equation}
	\mathcal{L} = 1 - \dfrac{1}{\abs{\mathcal{B}}} \sum_{\mathbf{h} \in \mathcal{B}} \dfrac{\abs{\mathbf{v}^\herm \mathbf{h}}^2}{\norm{\mathbf{h}}{2}^2},
	\label{eq:heuristic_loss}
\end{equation}
where $\mathcal{B}$ denotes the batch channel set and $\mathbf{v} = \left(\mathcal{D}_1\circ \mathcal{E}\right)\left(\mathbf{h}\right)$. One can show that Eq.~\eqref{eq:heuristic_loss} is minimized when the precoders are set to $\mathbf{v}^\star = \mathbf{h}\mathrm{e}^{\mathrm{j}\phi}/\norm{\mathbf{h}}{2}$. Thus, under the constraint of effective loss convergence, the trained decoder $\mathcal{D}_1$ provides the normalized channel affected by a phase shift. Through denormalization, one can then access a close approximation of the channel $\mathbf{h}$. Such channel can then be concatenated into a matrix, and used to form a precoding matrix, following the classical formulas of linear precoders. Note that, when considering this heuristic approach, the final decoding stage $\mathcal{D}_2$ does not present any learnable parameters so that the learning is done on the parameters of the $\mathcal{D}_1$ decoding stage.

\begin{algorithm}[t]
	\caption{Similarity subsampling~\cite{Taner2023}}
	\label{algo:sim_subsampling}
	\begin{algorithmic}[1]
	\Require Let $\tilde{\mathbf{D}} \in \mathbb{C}^{D \times \tilde{N}}$, $\tilde{\mathbf{Z}} \in \mathbb{R}^{d \times \tilde{N}}$, $\mathbf{D}^c \in \mathbb{C}^{D \times \left(N_c-\tilde{N}\right)}$, $\mathbf{Z}^c \in \mathbb{R}^{d \times \left(N_c-\tilde{N}\right)}$, s.t. $\mathbf{D} = \tilde{\mathbf{D}} \cap \mathbf{D}^c$, $\mathbf{Z} = \tilde{\mathbf{Z}} \cap \mathbf{Z}^c$.
	\ForEach {$\mathbf{h}_i \in \mathbf{D}^c$}
		\State $s \leftarrow \max\limits_{\mathbf{h}_j \in \tilde{\mathbf{D}}} \mathrm{sim}\left(\mathbf{h}_i, \mathbf{h}_j\right)$
		\State $\left\{k,l\right\} \leftarrow \argmax\limits_{\substack{\left(\mathbf{h}_m, \mathbf{h}_n\right) \in \tilde{\mathbf{D}}\\ 1 \leq m\neq n \leq \tilde{N}}} \mathrm{sim}\left(\mathbf{h}_m,\mathbf{h}_n\right)$
		\If{$s < \mathrm{sim}\left(\mathbf{h}_k,\mathbf{h}_l\right)$}
			\State With proba. $p: r\leftarrow k$, otherwise $r\leftarrow l$
			\State $\tilde{\mathbf{D}} \leftarrow \left(\tilde{\mathbf{D}} \setminus \mathbf{h}_r\right)\cup \mathbf{h}_i$
			\State $\tilde{\mathbf{Z}} \leftarrow \left(\tilde{\mathbf{Z}} \setminus \mathbf{z}_r\right)\cup \mathbf{z}_i$
		\EndIf
	\EndFor
	\Ensure $\tilde{\mathbf{D}}$ (Subs. channels), $\tilde{\mathbf{Z}}$ (Subs. chart locations)
	\end{algorithmic}
\end{algorithm}

\noindent\textbf{Parameter optimization.} The learnable parameters are conveyed by matrices $\mathbf{D} \in \mathbb{C}^{D \times N_c}$ and $\mathbf{Z} \in \mathbb{C}^{d \times N_c}$, frequency matrix $\mathbf{F} \in \mathbb{C}^{F \times d}$, in addition to the weights and biases of the decoder's MLP. As $d \ll D$, the main parameter complexity comes from the encoder. Indeed, $N_c$ needs to be high in order to have a good initialization in the \texttt{ISOMAP} algorithm, resulting in a large matrix $\mathbf{D}$. Optimizing the encoder parameter complexity is of paramount importance as the encoding stage is done at the UE level. Once trained, the BS needs to send the learned encoder parameters to the UEs: reducing the number of parameters in the encoder is then equivalent to reducing the encoder update overhead.

In order to reduce this parameter complexity, it is proposed to introduce an intelligent subsampling of matrices $\mathbf{D}$ and $\mathbf{Z}$.
Firstly, $\mathbf{D}$ and $\mathbf{Z}$ are initialized using \texttt{ISOMAP} as aforementioned. Then the similarity subsampling algorithm proposed in~\cite{Taner2023}, and recalled in Algorithm~\ref{algo:sim_subsampling}, is applied. Its objective is to discard channels that produce chart locations that are too similar. Indeed, similar chart locations can be considered as redundant, thereby providing no additional useful information. Theoretically, removing this redundancy from the calibration dataset allows to reduce the number of learnable parameters without impacting the global performance. More formally, Algorithm~\ref{algo:sim_subsampling} aims to solve the following minimax problem:
\begin{equation}
	\min \max\limits_{\left(\mathbf{h}_i, \mathbf{h}_j\right) \in \tilde{\mathbf{D}}} \mathrm{sim}\left(\mathbf{h}_i, \mathbf{h}_j\right),
\end{equation}
where $\tilde{\mathbf{D}}$ is the subsampled channel matrix and $\mathrm{sim}\left(\mathbf{h}_i, \mathbf{h}_j\right)$ is a similarity metric defined as the cosine similarity of the embedded channels:
\begin{equation}
	\mathrm{sim}\left(\mathbf{h}_i, \mathbf{h}_j\right) = \dfrac{\abs{\mathcal{E}\left(\mathbf{h}_i\right)^\herm \mathcal{E}\left(\mathbf{h}_j\right)}}{\norm{\mathcal{E}\left(\mathbf{h}_i\right)}{2}\norm{\mathcal{E}\left(\mathbf{h}_j\right)}{2}}.
\end{equation}
Note that, when using Algorithm~\ref{algo:sim_subsampling}, the maximum similarity in $\tilde{\mathbf{D}}$ is not hard-thresholded but rather indirectly controlled by the hyperparameter $\tilde{N}$.

\section{Experiments}\label{sec:experiments}

\noindent\textbf{Simulation settings.} Two separate datasets are considered: a \texttt{DeepMIMO}~\cite{deepmimo} dataset with the urban outdoor 'O1' scenario where the downlink central frequency is set to $3.5$GHz, and a \texttt{Sionna}~\cite{sionna} dataset generated with the \textit{Etoile} scene in Paris at $28$GHz. Both datasets considers $N_a = 64$ antennas and $N_s = 16$ subcarriers over a $20$ MHz bandwidth, $N_c = 5$k calibration channels, $3$k training channels and $2$k test channels.

Note that, for all following simulations, the precoders are learned for only one subcarrier, e.g. the central subcarrier. Although it is possible to extend this approach to all subcarriers, this will be addressed in future work.

\noindent\textbf{Hyperparameters and baselines.} When considering encoder parameter subsampling, the number of kept channels is set to $\tilde{N} = 10$ unless stated otherwise. Similarly, the embedding space dimension is set to $d=2$. For the decoder, $F=200$ frequencies are considered, and $\sigma_B$ is optimized for each dataset. Finally, the MLP layer size is set to $T=128$.

Baselines consist of an MLP autoencoder where the embedding layer has dimension $d$ as well as variants between the proposed approach and this MLP autoencoder: MLP encoder/RFF decoder and CC encoder/MLP decoder. Additionally, the RFF decoder with true locations as inputs is considered.

\begin{figure}[t]
	\centering
	\includegraphics[scale=.6]{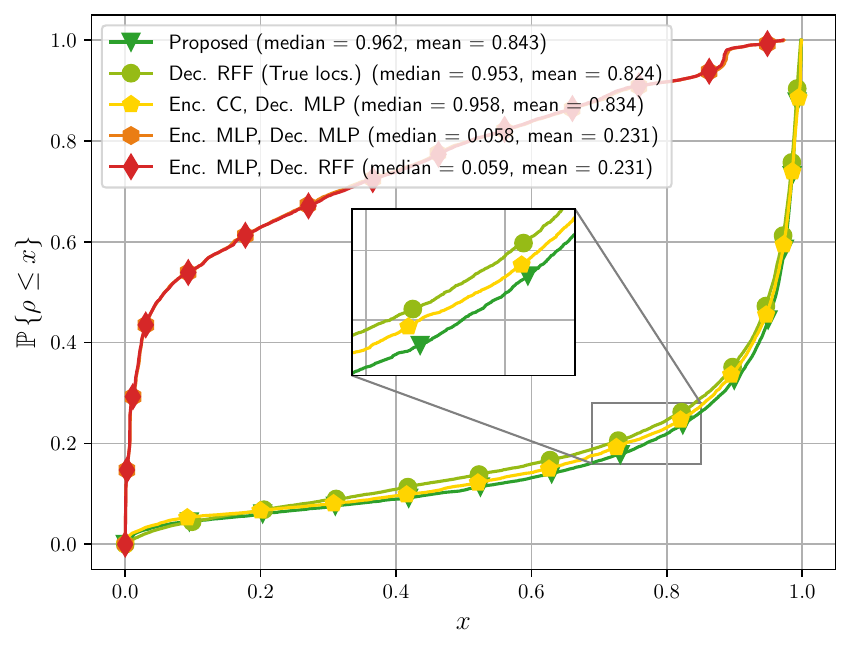}
	\caption{CDF of the squared cosine similarities: \texttt{Sionna} dataset}
	\label{fig:cdf_baselines}
\end{figure}

\noindent\textbf{Performance against baselines.} It clearly appears in Fig.~\ref{fig:cdf_baselines}, that CC, as a model-based encoder, has a significant role in the mono-UE model performance: every model with a MLP encoder presents poor performance. Additionally, it can be seen that the RFF decoder architecture provides a slight increase in performance. As the location manifold is inherently bidimensional, it is not surprising that bidimensional chart locations perform well in the proposed approach. This holds true under the continuity hypothesis of the underlying location-to-channel mapping~\cite{loc2chan_icassp,chatelier2024modelbasedlearningmultiantennamultifrequency}. Additionally, note that the learned bidimensional channel charts carry more meaningful informations than true locations, as the proposed architecture outperforms the RFF decoder used with true locations.

\begin{figure}[t]
	\centering
	\includegraphics[width=\columnwidth]{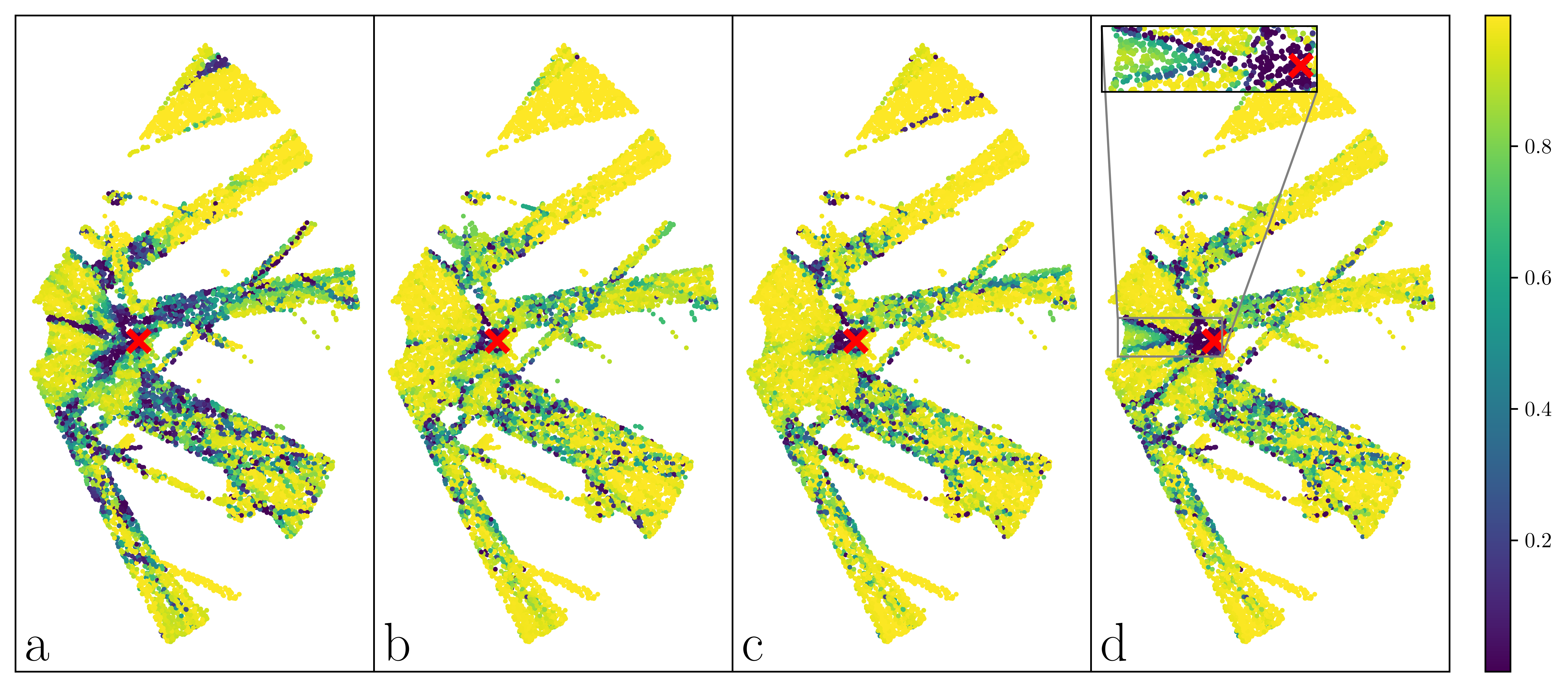}
	\caption{Squared cosine similarity maps: \texttt{Sionna} dataset}
	\label{fig:corr_map_enc_learning_impact}
\end{figure}

\noindent\textbf{Encoder learning and subsampling impact.} The different networks are denoted as follows: $\mathrm{a}$. No enc. learning, no subsampling, $\mathrm{b}$. Enc. learning, no subsampling, $\mathrm{c}$. Enc. learning, subsampling ($\tilde{N} = 100$), $\mathrm{d}$. Enc. learning, subsampling ($\tilde{N} = 10$). It is shown in Table~\ref{table:enc_learning_impact} and Fig.~\ref{fig:corr_map_enc_learning_impact} that allowing the encoder to be optimized during training is beneficial: both the median and mean normalized correlations are increased. Additionally, it is shown that the subsampling procedure does not introduce a significant performance drop. Conversely, when considering $\tilde{N} = 100$, a slight performance increase is observed. The minor performance drop when considering $\tilde{N} = 10$ can be attributed to the possibility of reduced scene coverage with such limited number of calibration channels. Indeed, as new chart locations are inferred from a convex combination of the calibration chart locations, considering too few calibration channels may result in insufficient coverage for some regions of the chart location space. This can be seen in the region immediately on the left of the BS (denoted by a red cross) in Fig.~\ref{fig:corr_map_enc_learning_impact}. Notwithstanding, with the considered hyperparameters, the subsampling approach with $\tilde{N} = 10$ allows almost a ninetyfold learnable parameter reduction with almost no performance drop.

% \begin{table}[t]
% 	\centering
% 		\begin{tabular}{lcccc}
% 			\toprule
% 			% \texttt{DeepMIMO} &  $\mathrm{a}$ & $\mathrm{b}$ & $\mathrm{c}$ & $\mathrm{d}$\\
% 			Network &  $\mathrm{a}$ & $\mathrm{b}$ & $\mathrm{c}$ & $\mathrm{d}$\\
% 			\toprule
% 			Params. & $101$k  & $10.3$M  & $306$k & $121$k \\
% 			\toprule
% 			\texttt{DeepMIMO} \\
% 			\toprule
% 			Median $\rho$ &   $0.88$ &  $\mathbf{0.94}$ &  $\mathbf{0.94}$ & $\mathbf{0.94}$ \\
% 			\midrule 
% 			Mean $\rho$ &   $0.79$ &  $0.86$ &  $\mathbf{0.88}$ & $\mathbf{0.88}$ \\
% 			\toprule
% 			\texttt{Sionna} \\
% 			\toprule
% 			Median $\rho$ &   $0.91$ &  $0.96$ &  $\mathbf{0.97}$ &  $0.96$ \\
% 			\midrule
% 			Mean $\rho$ &   $0.73$ &  $0.85$ & $\mathbf{0.87}$ &  $0.84$ \\
% 			\bottomrule
% 		\end{tabular}
% 	\caption{Mono-UE precoder regression performance}
% 	\label{table:enc_learning_impact}
% \end{table}

\begin{table}[t]
	\centering
	% \scriptsize
		\begin{tabular}{clccccc}
			\toprule
			& Network &  $\mathrm{a}$ &  $\mathrm{b}$ &  $\mathrm{c}$ &  $\mathrm{d}$ \\
            \midrule
            & Params. &  $101$k  & $10.3$M  & $306$k & $121$k \\
            \toprule
			\multirow[c]{2}{*}{\vspace{-2mm} \texttt{Sionna}}& Median $\rho$ & $0.88$ &  $\mathbf{0.94}$ &  $\mathbf{0.94}$ & $\mathbf{0.94}$ \\
			\cmidrule(l){2-6}& Mean $\rho$ & $0.79$ &  $0.86$ &  $\mathbf{0.88}$ & $\mathbf{0.88}$ \\
			\midrule
			\multirow[c]{2}{*}{\vspace{-2mm} \texttt{DeepMIMO}}& Median $\rho$ & $0.91$ &  $0.96$ &  $\mathbf{0.97}$ &  $0.96$ \\
			\cmidrule(l){2-6}& Mean $\rho$ &  $0.73$ &  $0.85$ & $\mathbf{0.87}$ &  $0.84$ \\
			\bottomrule
		\end{tabular}
		\caption{Mono-UE precoder regression performance}
		\label{table:enc_learning_impact}
\end{table}

% \vspace{-2\baselineskip}

\noindent\textbf{Compression ratio impact.} The compression ratio can be defined as:
\begin{equation}
	\gamma = \dfrac{2 N_a N_s}{d+1},
\end{equation}
as downlink channel vectors are composed of $N_a N_s$ complex coefficients. The $+1$ term at the denominator arises from the requirement of channel norm transfer to the BS, as a denormalization step is performed.

\begin{figure}[t]
	\centering
	\includegraphics[width=.96\columnwidth]{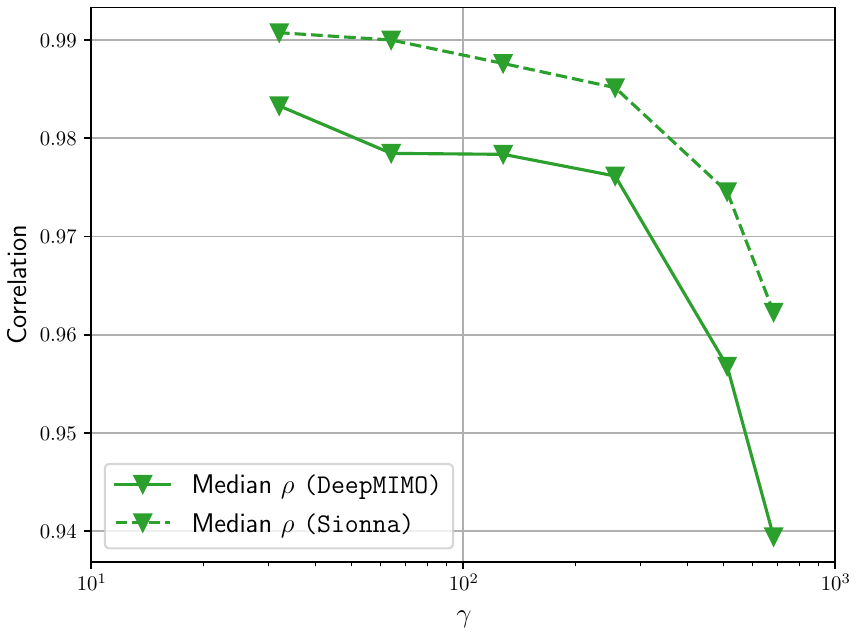}
	\caption{Median $\rho$ evolution with the compression ratio}
	\label{fig:compression_ratio_evolution}
\end{figure}

Fig.~\ref{fig:compression_ratio_evolution} presents the evolution of the median squared cosine similarity of the proposed approach with respect to the compression ratio. This is achieved by varying the embedding space dimension $d$. It can be seen that increasing $d$ slightly increases performance. It should also be noted that it is possible to reach good performance with $d=2 \Rightarrow \gamma \simeq 1024$, which significantly exceeds compression ratios reached with classical autoencoders (around $64$). This confirms the interest of the proposed model-based approach.

\begin{figure}[t]
	\centering
	\includegraphics[width=.96\columnwidth]{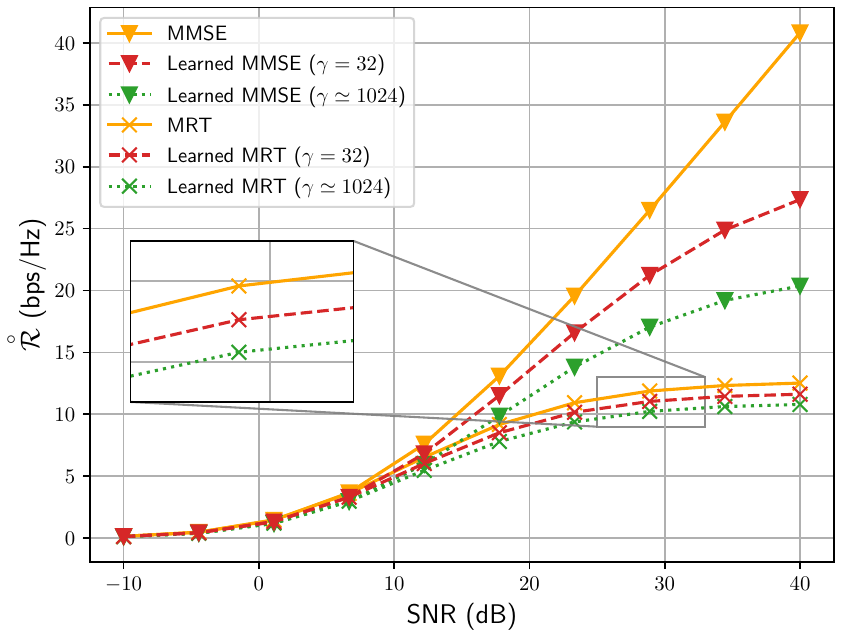}
	\caption{SR performance: \texttt{Sionna} dataset}
	\label{fig:ergodic_sr}
\end{figure}

\noindent\textbf{Muti-UE beamforming performance.} It can be seen in Fig.~\ref{fig:ergodic_sr} that the learned MRT precoding matrix exhibits the same behavior as the actual MRT precoder. Indeed, the learned MRT can be seen as the concatenation of mono-UE MRT precoders affected by a phase shift. Such phase shift introduces the observed performance degradation. Additionally, it can be seen that, at high SNR, the learned MMSE precoder, obtained by considering the inferred precoding matrix $\mathbf{W}$ as a channel matrix, outperforms the actual MRT precoder. This outlines the potential of the proposed approach, as it is possible to perform inter-UE interference cancellation with the proposed heuristic learning procedure that only considers individual UEs and compressed channel representations. Additionally, considering a lower compression ratio increases the mono-UE precoder learning performance, as observed in Fig.~\ref{fig:compression_ratio_evolution}, which translates into the multi-UE setting with better inter-UE interference cancellation. 

\section{Conclusion and perspectives}\label{sec:conclusion}

In this paper, the problem of task-oriented CSI compression was studied. A model-based neural architecture, composed of a CC encoder and a RFF decoder, was proposed with the objective of maximizing the SR in a MU-MIMO system. An end-to-end heuristic learning strategy was introduced, and the number of learnable parameters was optimized through a subsampling method. The proposed method performance have been evaluated on realistic data, showing its superior performance against several baselines at high compression ratios. Additionally, this study outlined the performance of CC as a channel compression method for other tasks than localization.
Future work will explore SR optimization through the introduction of a UE grouping partition policy and a learnable power allocation strategy.

\bibliographystyle{./IEEEtran.bst}
\bibliography{./refs/paper_biblio.bib}

\end{document}